\pdfoutput=1
\PassOptionsToPackage{unicode}{hyperref}
\PassOptionsToPackage{hyphens}{url}
\PassOptionsToPackage{dvipsnames,svgnames,x11names}{xcolor}
\documentclass[
  11pt,
  letterpaper,
]{article}
\usepackage{xcolor}
\usepackage[margin=2.5cm]{geometry}
\usepackage{amsmath,amssymb}
\usepackage{iftex}
\ifPDFTeX
  \usepackage[T1]{fontenc}
  \usepackage[utf8]{inputenc}
  \usepackage{textcomp} 
\else 
  \usepackage{unicode-math} 
  \defaultfontfeatures{Scale=MatchLowercase}
  \defaultfontfeatures[\rmfamily]{Ligatures=TeX,Scale=1}
\fi
\usepackage{lmodern}
\ifPDFTeX\else
\fi
\IfFileExists{upquote.sty}{\usepackage{upquote}}{}
\IfFileExists{microtype.sty}{
  \usepackage[]{microtype}
  \UseMicrotypeSet[protrusion]{basicmath} 
}{}
\makeatletter
\@ifundefined{KOMAClassName}{
  \IfFileExists{parskip.sty}{%
    \usepackage{parskip}
  }{
    \setlength{\parindent}{0pt}
    \setlength{\parskip}{6pt plus 2pt minus 1pt}}
}{
  \KOMAoptions{parskip=half}}
\makeatother
\usepackage{graphicx}
\makeatletter
\newsavebox\pandoc@box
\newcommand*\pandocbounded[1]{
  \sbox\pandoc@box{#1}%
  \Gscale@div\@tempa{\textheight}{\dimexpr\ht\pandoc@box+\dp\pandoc@box\relax}%
  \Gscale@div\@tempb{\linewidth}{\wd\pandoc@box}%
  \ifdim\@tempb\p@<\@tempa\p@\let\@tempa\@tempb\fi
  \ifdim\@tempa\p@<\p@\scalebox{\@tempa}{\usebox\pandoc@box}%
  \else\usebox{\pandoc@box}%
  \fi%
}
\def\fps@figure{htbp}
\makeatother
\NewDocumentCommand\citeproctext{}{}
\NewDocumentCommand\citeproc{mm}{%
  \begingroup\def\citeproctext{#2}\cite{#1}\endgroup}
\makeatletter
 \let\@cite@ofmt\@firstofone
 \def\@biblabel#1{}
 \def\@cite#1#2{{#1\if@tempswa , #2\fi}}
\makeatother
\newlength{\cslhangindent}
\newlength{\csllabelwidth}
\newenvironment{CSLReferences}[2] 
 {\begin{list}{}{%
  \setlength{\itemindent}{0pt}
  \setlength{\leftmargin}{0pt}
  \setlength{\parsep}{0pt}
  \ifodd #1
   \setlength{\leftmargin}{\cslhangindent}
   \setlength{\itemindent}{-1\cslhangindent}
  \fi
  \setlength{\itemsep}{#2\baselineskip}}}
 {\end{list}}
\usepackage{calc}

\newcommand{\CSLLeftMargin}[1]{\parbox[t]{\csllabelwidth}{\strut#1\strut}}
\newcommand{\CSLRightInline}[1]{\parbox[t]{\linewidth - \csllabelwidth}{\strut#1\strut}}

\providecommand{\tightlist}{%
  \setlength{\itemsep}{0pt}\setlength{\parskip}{0pt}}
\usepackage{float}
\usepackage{bookmark}
\IfFileExists{xurl.sty}{\usepackage{xurl}}{} 
\hypersetup{
  pdftitle={Mode-tunable inter-core coupling of photon-number-resolved quantum light in a telecom multicore fiber},
  colorlinks=true,
  linkcolor={Maroon},
  filecolor={Maroon},
  citecolor={Blue},
  urlcolor={Blue},
  pdfcreator={LaTeX via pandoc}}

\title{Mode-tunable inter-core coupling of photon-number-resolved
quantum light in a telecom multicore fiber}
\author{}
\date{}

\begin{document}
\maketitle

\textbf{Authors:} Chiran Wijesundara\(^{1}\), Yasmin Sarhan\(^{1}\),
Henry Strong\(^{1}\), Jeffrey Cai\(^{1}\), Dan T. Nguyen\(^{2}\), Tim
Thomay\(^{1,\dagger}\)\\
\(^{1}\)\emph{Department of Physics, University at Buffalo, The State
University of New York, Buffalo, NY 14260, USA}\\
\(^{2}\)\emph{Science and Technology Division, Corning Research and
Development Corporation, Corning, NY 14831, USA}\\
\(^{\dagger}\)\emph{Corresponding author: thomay@buffalo.edu}

\begin{center}\rule{0.5\linewidth}{0.5pt}\end{center}

\subsection{Abstract}\label{abstract}

Multicore optical fibers (MCFs), originally developed to scale classical
telecommunications, are rapidly gaining traction as a quantum platform
that carries high-dimensional key distribution, entanglement, and
photonic quantum walks. Yet quantum light has occupied these fibers only
as single-mode path channels; the spatial-mode structure within each
core has remained untouched. Here we operate a telecom-grade 39-core
fiber far from its design wavelength, where each core supports exactly
two LP mode groups (LP\(_{01}\), LP\(_{11}\)), and measure heralded
photon-number states \(|1\rangle\) and \(|2\rangle\), resolving the
photon number simultaneously with the intra-core spatial mode in one
detection configuration and with the core index in the other. Displacing
the launch prepares a coherent LP\(_{01}\)--LP\(_{11}\) superposition
whose single-sided intensity pattern demonstrates that the intermodal
phase is preserved over the fiber: once the detector alignment is
constrained by a drift-resolved joint fit, the measured asymmetry (0.4\%
vs 47\% on opposing detector pixels) is incompatible with an incoherent
mode mixture. Exploiting the constraint that only two modes exist, we
back-calculate the launch trajectory from the photon data alone, linking
modal content, launch coupling efficiency, and launch displacement.
Because the two modes couple to neighboring cores at different rates
(\(\kappa_{\mathrm{LP11}}/\kappa_{\mathrm{LP01}}\) \(\approx\) 2.4), the
launched mode content acts as a state-selectable coupling knob. The
effective inter-core coupling of the passive fiber is set by the launch
alone (measured 3.9--13.1\%), a tunability that otherwise requires
reconfiguring the lattice with electro-optic, thermo-optic, or
mechanical elements, while the preserved intermodal phase adds a second,
interferometric control axis. The photon-number classes are certified by
the herald statistics, timing, and artifact-vetoed pair distributions of
more than \(10^{10}\) time-tagged events, and the propagation is
photon-number-independent. The per-photon inter-core crosstalk of
heralded \(|2\rangle\) states equals that of \(|1\rangle\) within
errors, bounding a Fock-state dependence of the per-photon crosstalk to
\(|\Delta x/x|\) \textless{} 6.1\% at 95\% confidence level (CL), and
the imaged mode distributions of \(|1\rangle\) and \(|2\rangle\) agree
to better than 1\%. This measured linear-optics null, validated against
exact Fock-space simulation (Perceval, The Walrus, and closed form,
agreeing to \(10^{-13}\)), establishes the reference baseline against
which genuine multiphoton interference in fiber-based quantum walks must
be certified; in addition, we show quantitatively how coincidence-based
event selection produces \emph{spurious} photon-number-dependent
crosstalk if accidentals are not subtracted, a pitfall for multiphoton
experiments in coupled-waveguide systems. The results establish
off-design operation of deployed fiber as a calibrated, phase-preserving
platform for high-dimensional quantum encoding with photon-number and
spatial-mode degrees of freedom.

\subsection{1. Introduction}\label{introduction}

Optical fibers are approaching their classical capacity limit, and
space-division multiplexing, that is, parallel transmission through the
cores of a multicore fiber (MCF) or the transverse modes of a few-mode
fiber, was developed to overcome
it\textsuperscript{\citeproc{ref-Richardson2013}{1}--\citeproc{ref-Awaji2019}{3}}.
These spatial channels have, however, also found an exciting second life
in quantum technology\textsuperscript{\citeproc{ref-Xavier2020}{4}}:
multicore fibers have hosted photonic quantum walks across their coupled
cores\textsuperscript{\citeproc{ref-Nguyen2020}{5}}, high-dimensional
quantum key
distribution\textsuperscript{\citeproc{ref-Canas2017}{6},\citeproc{ref-Ding2017}{7}},
path-encoded qudits over kilometer-scale
links\textsuperscript{\citeproc{ref-DaLio2021}{8}}, four-dimensional
spatial entanglement\textsuperscript{\citeproc{ref-Lee2017}{9}},
multidimensional entangled-state
generation\textsuperscript{\citeproc{ref-Gomez2021}{10}}, and, recently,
hyper-entanglement in three degrees of
freedom\textsuperscript{\citeproc{ref-Achatz2023}{11}} and
space-division-multiplexed entanglement-based cryptography over 19
cores\textsuperscript{\citeproc{ref-Ortega2024}{12}}. This progress
rests on a practical advantage that has been measured directly: the
relative phase between MCF cores drifts on second rather than
millisecond timescales compared with separate fibers over a 2 km
link\textsuperscript{\citeproc{ref-DaLio2020}{13}}, which is what makes
MCF-integrated interferometers and multiport splitters
viable\textsuperscript{\citeproc{ref-Carine2020}{14}}.

In all of these experiments, however, quantum information occupies only
the \emph{which-core} degree of freedom: the cores are operated at their
design wavelength, each supporting a single mode. The spatial structure
available \emph{within} each core has remained untouched; in fact, when
telecom fiber is operated below its cutoff wavelength, where
higher-order modes appear, those modes have historically been filtered
away as a nuisance\textsuperscript{\citeproc{ref-MeyerScott2010}{15}}.
Mode-resolved quantum optics has instead developed in single-core
geometries: photon pairs generated and sorted in the LP-mode basis of a
few-mode
fiber\textsuperscript{\citeproc{ref-Sulimany2022}{16},\citeproc{ref-Cui2019}{17}},
entangled-pair distribution over a two-mode
fiber\textsuperscript{\citeproc{ref-Cui2017}{18}}, spatially entangled
qubits through hollow-core photonic-crystal
fiber\textsuperscript{\citeproc{ref-Loffler2011}{19}},
orbital-angular-momentum qudit transmission and entanglement
distribution through specialty
fibers\textsuperscript{\citeproc{ref-Cozzolino2019}{20},\citeproc{ref-Cao2020}{21}},
two-photon speckle in a continuum multimode
fiber\textsuperscript{\citeproc{ref-Defienne2016}{22}}, all-fiber
shaping of single photons and entangled-pair correlations through
multimode fiber by controlled mechanical
perturbations\textsuperscript{\citeproc{ref-Shekel2023}{23}},
spatial-mode shaping of SPDC photon-pair light in a few-mode fiber by
fiber-piano perturbation and launch
misalignment\textsuperscript{\citeproc{ref-Wu2024}{24}}, and most
recently the spatial-mode-multiplexed co-propagation of quantum and
classical signals in few-mode
fiber\textsuperscript{\citeproc{ref-Zia2024}{25}}. None of these
platforms combines intra-core modes with a core lattice, and none
resolves the photon number of the guided light beyond post-selected pair
coincidences.

A lattice of coupled waveguides is also the canonical host of photonic
quantum walks: quantum-walk dynamics were first observed with classical
light in waveguide
lattices\textsuperscript{\citeproc{ref-Perets2008}{26}}, the two-photon
correlation theory and its classical emulation
followed\textsuperscript{\citeproc{ref-Bromberg2009}{27}}, and
correlated-photon walks were then realized in integrated waveguide
arrays\textsuperscript{\citeproc{ref-Peruzzo2010}{28}--\citeproc{ref-Sansoni2012}{30}},
underpinning boson
sampling\textsuperscript{\citeproc{ref-Spring2013}{31}};
see\textsuperscript{\citeproc{ref-Grafe2016}{32},\citeproc{ref-Qiang2024}{33}}
for reviews. Making such lattices \emph{tunable} currently means
reconfiguring the lattice itself, whether through per-site thermo-optic
phase shifters and programmable interferometer
meshes\textsuperscript{\citeproc{ref-Carolan2015}{34},\citeproc{ref-Wang2018}{35}}
or through mechanical reconfiguration of birefringent
elements\textsuperscript{\citeproc{ref-Fenwick2024}{36}}. Multicore
fiber brings the complementary virtues of kilometer-scale low-loss
propagation, and quantum-walk dynamics have been demonstrated in the
very fiber used here, although with classical light emulating the
single-photon regime\textsuperscript{\citeproc{ref-Nguyen2020}{5}}.

Photon number is the remaining, largely untapped resource. While
time-frequency encodings reach Hilbert-space dimensionalities in the
hundreds\textsuperscript{\citeproc{ref-Chang2021}{37}}, with chip-scale
microcombs as a scalable frequency-bin
platform\textsuperscript{\citeproc{ref-Kues2019}{38}}, the photon-number
axis \(|n\rangle\) of a single spatial mode adds an alphabet that no
lossless linear network can erase. The measurement toolkit for it has
matured, ranging from number-resolving detector
cascades\textsuperscript{\citeproc{ref-Thomay2017}{39},\citeproc{ref-Eisaman2011}{40}}
and machine-learned state
classification\textsuperscript{\citeproc{ref-Xu2024}{41}} to
photon-number-resolving detectors in the adjacent 850--950 nm
band\textsuperscript{\citeproc{ref-Los2024}{42}}. What has been missing
is a fiber platform that combines the photon-number axis with a scalable
spatial-mode structure.

Here we simultaneously resolve, for the first time to our knowledge, the
photon-number and intra-core spatial-mode degrees of freedom of quantum
light in a multicore fiber. Operating the 39-core fiber of
Ref.\textsuperscript{\citeproc{ref-Nguyen2020}{5}} at 800 nm, where each
core supports exactly two LP mode groups, we obtain four results. First,
a launch displacement prepares a coherent LP\(_{01}\)--LP\(_{11}\)
superposition whose relative phase survives the fiber: each core carries
a spatial qubit, not a classical mixture. Second, because the two modes
couple to neighboring cores at different rates, the launch selects the
effective inter-core coupling of the passive lattice over a factor of
three (3.9--13.1\%). This is a single global knob rather than per-site
programmability, but it requires no electrodes, heaters, or
interferometer meshes in the lattice. Third, the propagation is measured
to be photon-number-independent, a result we term the ``linear-optics
null'', with a Fock-state dependence of the per-photon crosstalk bounded
below 6.1\% (95\% CL, confidence level); we further show how easily
coincidence-filtered analyses violate this null spuriously, and validate
the corrections against exact multiphoton simulation. Fourth, the
two-mode constraint turns the photon data into its own diagnostic: we
back-calculate the launch trajectory, its drift, and the intermodal
phase \(\delta\beta L\) from the heralded images alone.

\subsection{2. Concept and experimental
setup}\label{concept-and-experimental-setup}

\includegraphics[width=0.95\linewidth,height=\textheight,keepaspectratio]{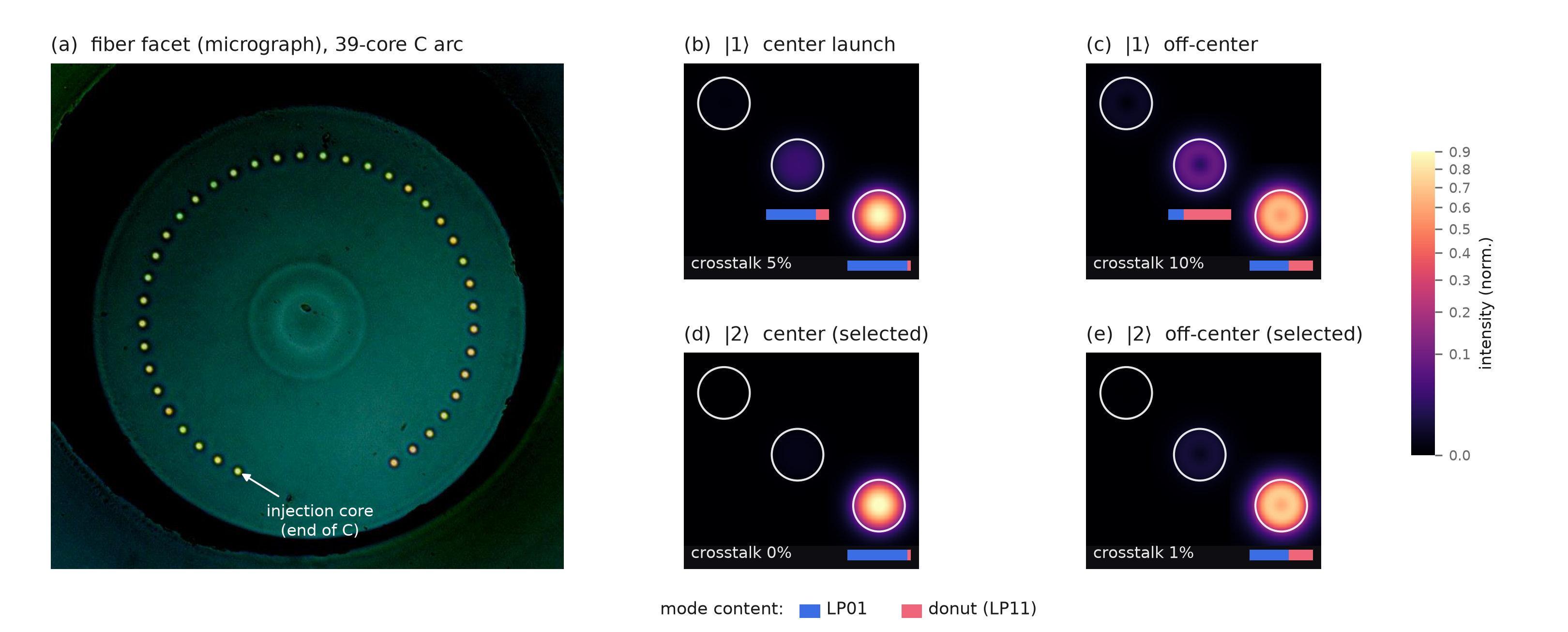}~

\textbf{Fig. 1 \textbar{} The multicore ring fiber and the coupled-mode
model.} (a) Micrograph of the facet of the 39-core ring fiber (image
courtesy of Corning Incorporated): the gap of several core positions
opens the ring into a C-shaped arc and makes the core chain a
one-dimensional lattice; light is injected into the marked core at one
end of the arc. (b,c) Calibrated coupled-mode model of the output
intensity in the launch core and its two detected neighbors for heralded
single photons with centered (b) and displaced (c) launch; bars give the
modal content (LP01 vs LP11) and the per-photon crosstalk. (d,e) The
same for heralded two-photon events when coincidence-based event
selection is applied: the apparent crosstalk collapses (10\% to 1\% for
the displaced launch) although the physics is unchanged, illustrating
the event-selection artifact quantified in Section 5. Panels (b-e) are
event-averaged model intensities; the single-shot coherent pattern is
shown in Fig. 4(d).

\includegraphics[width=0.95\linewidth,height=\textheight,keepaspectratio]{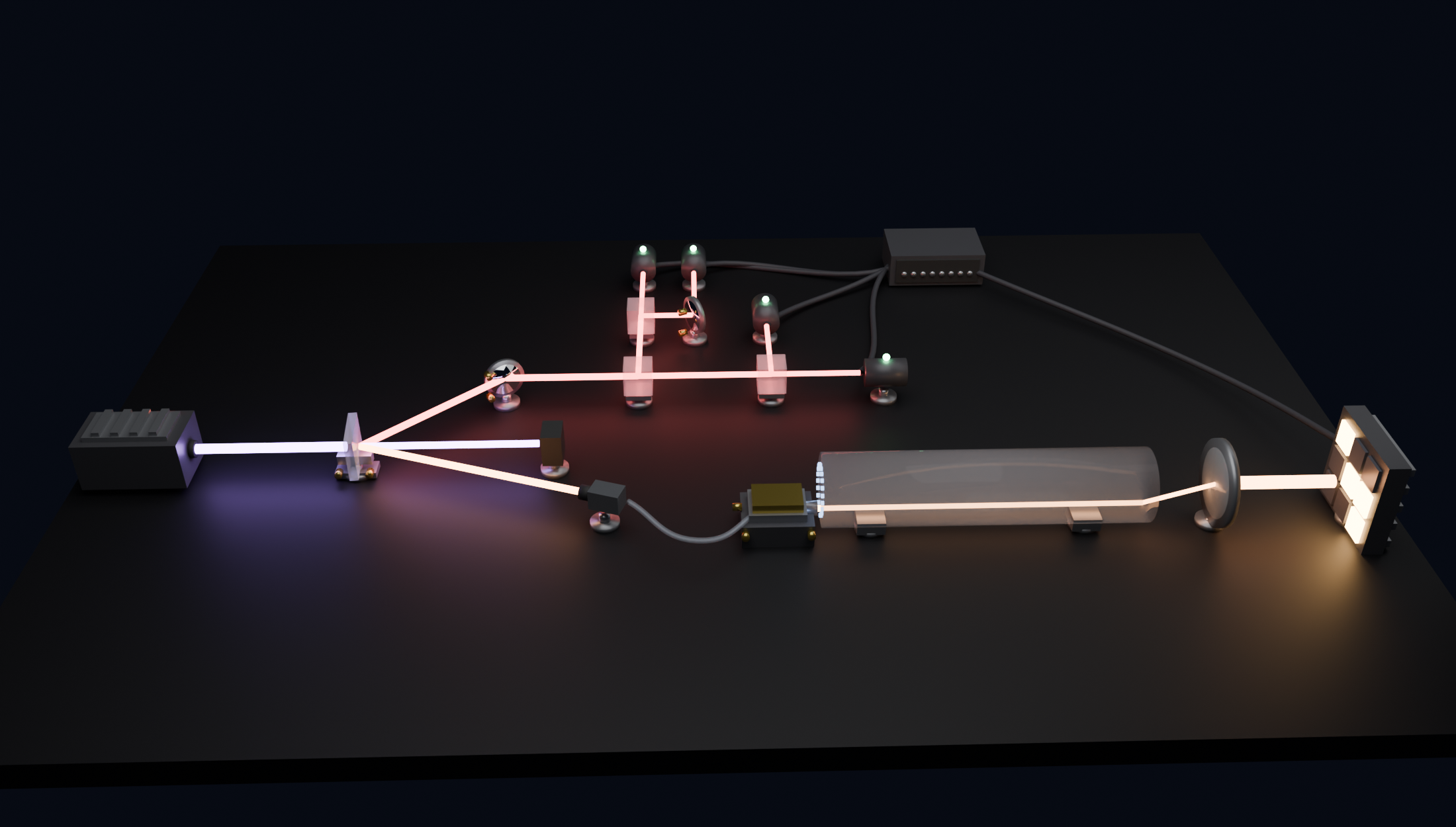}~

\textbf{Fig. 2 \textbar{} Experimental setup (3D render).} A
frequency-doubled Ti:Sapphire regenerative amplifier pumps type-II SPDC
in BBO. Idler photons pass a monochromator and a fiber-beam-splitter
cascade onto four APDs, which together form the photon-number-resolving
herald; signal photons are launched by a piezo-positioned single-mode
fiber into the end-of-arc core of the multicore fiber, whose output
facet is imaged onto the APD array (four read-out pixels). All eight
detector channels share one time base. In the core-resolved
configuration the pixels SE, C and NW+W collect cores 1, 2 and 3.

\includegraphics[width=0.95\linewidth,height=\textheight,keepaspectratio]{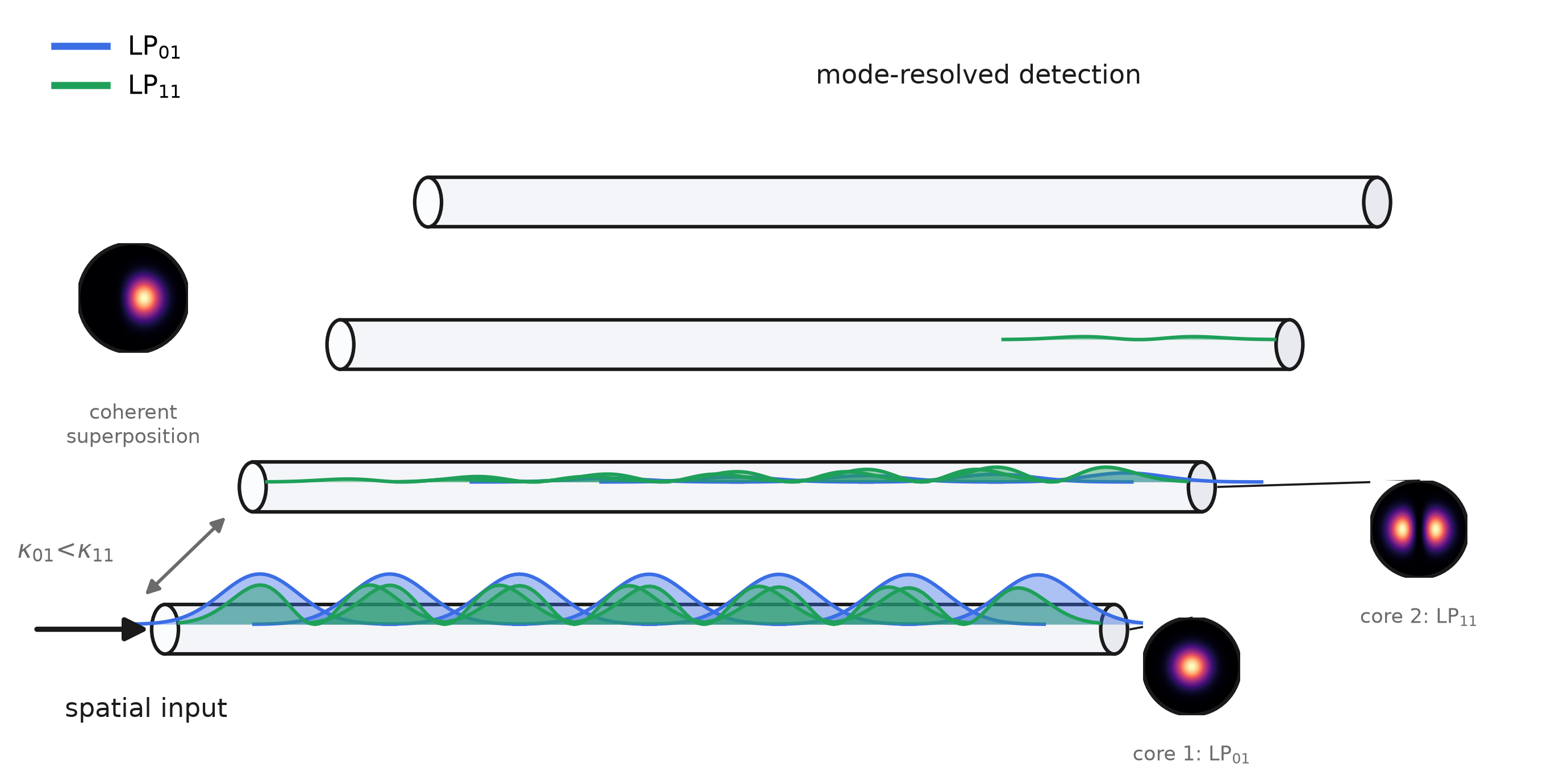}~

\textbf{Fig. 3 \textbar{} Spatial-mode coupling concept.} LP01 (blue)
and LP11 (green) amplitudes along neighboring cores of the array: the
launched mode content sets the strength of the evanescent inter-core
coupling (kappa01 \textless{} kappa11), and both the input and the
imaged output are mode-resolved. Bump heights follow the calibrated
coupled-mode powers over the sample; circular insets show the
facet-plane intensities (coherent superposition at the input; LP01 vs
LP11 at the output).

The fiber (Fig. 1) is the same Corning multicore ring fiber as in
Ref.\textsuperscript{\citeproc{ref-Nguyen2020}{5}}, to which we refer
for its design and fabrication: 39 identical cores on a ring, with a gap
of several core positions that opens the ring into a C-shaped arc. The
gap makes the core chain a genuine 1D lattice with two free ends; light
is injected into the first core at one end of the arc, so the walker
starts at the end of an open chain. At its design wavelength of 1550 nm
each core is single-mode, and ballistic quantum walks were demonstrated
in this fiber with classical light (localized walks in its
Fibonacci-spaced sibling
fiber)\textsuperscript{\citeproc{ref-Nguyen2020}{5}}. At our operating
wavelength of 800 nm the launch core guides two LP mode groups,
LP\(_{01}\) and LP\(_{11}\), as the mode-resolved images of Section 3
show directly, turning each lattice site into a two-level spatial
system. All mode profiles below are those of a step-index core with
effective parameters \(a_{\mathrm{eff}}\) = 4.5 \(\mu\)m and
\(V_{\mathrm{eff}}\) = 3.57, chosen to reproduce the measured images,
and all inter-core couplings are fitted to the data (Section 4); no
nominal fiber parameters enter the analysis.

Photon pairs at \(\sim 800\) nm are generated by type-II SPDC (BBO)
(Fig. 2) pumped by a frequency-doubled Ti:Sapphire regenerative
amplifier (measured sync rate 249,733 Hz); the higher-order pair
emission of such pulsed BBO SPDC sources at 800 nm is characterized
in\textsuperscript{\citeproc{ref-Carvalho2025}{43}}. Signal photons are
butt-coupled into one core through a 780HP single-mode fiber on a piezo
stage; the transverse launch position s is the state-preparation
control. Idler photons pass a monochromator whose 50 \(\mu\)m exit slit
feeds a multimode fiber, then a cascaded fiber-beam-splitter arm with
four APDs, which together form a photon-number-resolving
herald\textsuperscript{\citeproc{ref-Thomay2017}{39},\citeproc{ref-Xu2024}{41}};
the narrow idler selection also acts as a nonlocal spectral filter on
the heralded signal (Section 3). The MCF output is imaged onto an APD
array (3\(\times 3\) grid, pitch 75 \(\mu\)m, pixel radius 15 \(\mu\)m)
of which four pixels are read out, in one of two optical configurations:
\textbf{(i) core-resolved}, where the magnification aligns three
neighboring cores with three pixels (crosstalk measurements); and
\textbf{(ii) mode-resolved}, where a higher magnification (fitted
11\(\times\)) spreads the image of the launch core across the array
(modal measurements). Because the core-resolved configuration samples
each core with a pixel, per-core collection efficiencies are not
identical; crosstalk comparisons \emph{within and between runs at fixed
alignment} are unaffected, and the photon-number comparison (Section 5)
is unaffected (efficiencies cancel between h1 and h2), but absolute
coupling constants carry a per-core collection systematic. All eight
channels share one MultiHarp time base (T3 mode; intra-pulse arrival
times, ``dtime'', binned at 320 ps); in total, more than \(10^{10}\)
time-tagged detection events were recorded across the measurement
series, and every analysis below acts on this one dataset (Data
availability).

\textbf{Conventions.} Herald multiplicities are written h1, h2 and the
corresponding heralded classes \(|1\rangle\), \(|2\rangle\); lossy
heralding makes m a lower bound on the photon number, and the
comparisons below are insensitive to this admixture. The four read-out
pixels are labeled NW, C, SE, W; in the core-resolved configuration SE
collects core 1 (the launch core), C core 2, and NW+W core 3 (Fig. 2).

\textbf{Coincidence analysis.} Photons are grouped by laser pulse (sync
index); the herald multiplicity m (number of distinct herald detectors
firing) prepares the heralded class \(|m\rangle\). Signal photons count
within a dtime window centered at the calibrated herald--signal offset
\(\Delta_0\) = 4.16 ns (arm path difference); results are
window-insensitive from 0.5--3 ns. Accidental coincidences, estimated by
matching heralds of pulse k against signals of pulse k+1, are subtracted
from every counter. This correction is not optional: the raw same-pulse
crosstalk (16\% centered / 32\% off-center) collapses to 4.6\%
(centered) and 4.7\% (off-center, low-coupling direction) after
subtraction, and Section 5 shows that omitting it manufactures a
spurious photon-number dependence.

\textbf{Quantum characterization of the heralded states.} The
photon-number classes are prepared and certified on the herald side.
Over a single 20 h centered run the four-detector cascade registered
\(1.16\times10^{9}\), 6.9\(\times10^{7}\), 2.7\(\times10^{6}\), and
5.1\(\times10^{4}\) pulses with herald multiplicity m = 1, 2, 3, 4; the
successive ratios (0.059, 0.039, 0.019) decrease as expected for a lossy
four-detector cascade sampling the multi-pair emission of a pulsed SPDC
source\textsuperscript{\citeproc{ref-Thomay2017}{39},\citeproc{ref-Xu2024}{41},\citeproc{ref-Carvalho2025}{43}},
and the m = 2 class alone contains \(\sim 7\times10^{7}\) heralds, so
every \(|2\rangle\) observable below is limited by systematics rather
than by statistics. The heralded signal is a prompt, herald-correlated
peak: all four signal channels share one arrival-time bin at
\(\Delta_0\) = 4.16 ns, and the net (accidental-subtracted) per-core
counts are constant to within statistics as the coincidence window is
varied from 0.5 to 3 ns, which is the signature of true heralded photons
on top of an uncorrelated background. The background is quantified by
the accidental estimate: in the raw same-pulse counters it accounts for
70--85\% of the apparent crosstalk, and after subtraction the per-photon
crosstalk of the \(|2\rangle\) class agrees with that of \(|1\rangle\)
within errors (Section 5). Genuine two-photon events are identified
directly: the \(\sim 2.3\times10^{3}\) heralded \(|2\rangle\) events
with both photons detected in different cores follow independent-photon
multinomial statistics, and the pair time-delay distribution (Methods,
Fig. 6) separates prompt genuine pairs (\textless{} 0.64 ns) from the
delayed APD-array optical-crosstalk bump (\(\sim 1\) ns) and from
afterpulses (\textgreater{} 13 ns), so that the two-photon sample used
in Sections 5 and 6 is artifact-vetoed rather than modeled. The
intermodal coherence of Section 3 (visibility \(V \geq 0.95\) for the
LP\(_{01}\)--LP\(_{11}\) cross term) is a state property of the same
heralded photons, and the exact Fock-space model of Section 5 (three
independent engines agreeing to 10\(^{-13}\)) supplies the reference
multiphoton statistics against which all of these observables are
compared.

\subsection{3. Coherent spatial-mode
control}\label{coherent-spatial-mode-control}

\includegraphics[width=0.95\linewidth,height=\textheight,keepaspectratio]{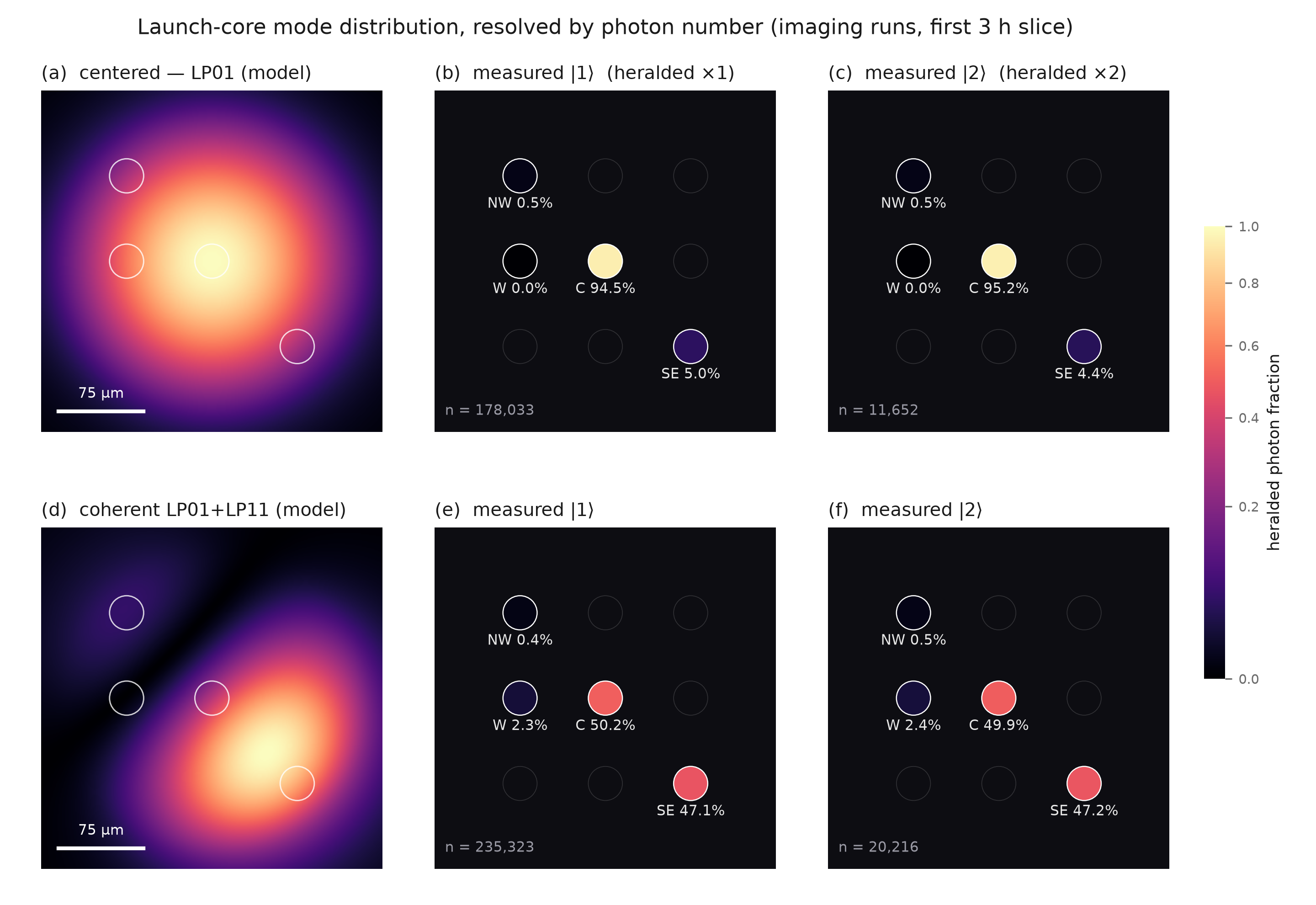}~

\textbf{Fig. 4 \textbar{} Measured heralded mode distributions.} (a)
Modelled LP01 intensity for the centered launch; (b,c) measured heralded
pixel distributions for the centered launch with photon number one and
two; (d) modelled intensity of the coherent LP01+LP11 superposition,
reproducing the observed single-sided lobe that an incoherent mixture
cannot produce; (e,f) measured distributions for the displaced launch
with photon number one and two. Photon numbers one and two agree to
better than 1\%; all fractions refer to the first of eight time slices.

With the array imaging the launch core, the heralded pixel distributions
directly display the mode (Fig. 4). A centered launch puts 94.5\% of the
heralded photons on the central pixel, identifying the LP\(_{01}\)
fundamental (the 24 h imaging runs are cut into eight 3 h time slices;
all pixel fractions are quoted for the first slice, and the launch drift
over the run is analysed below). A launch displaced by 2.2 \(\mu\)m
along the array diagonal redistributes the light into a single displaced
lobe (central pixel 50.2\%, diagonal pixel 47.1\%, opposite pixel
0.4\%).

Intermodal LP\(_{01}\)/LP\(_{11}\) interference is a well-established
classical effect: it underlies two-mode-fiber interferometry, and the
S\(^2\) mode-imaging technique uses exactly this spatial-spectral
interference to characterize few-mode
fibers\textsuperscript{\citeproc{ref-Nicholson2008}{44}}. What is new
here is that the pattern is resolved photon-by-photon on heralded Fock
states, and that its asymmetry is used quantitatively. The asymmetry of
this pattern is a measurement of the intermodal coherence. An
\emph{incoherent} mixture of LP\(_{01}\) with the LP\(_{11}\) doublet
(arbitrary weights, arbitrary lobe orientation, even arbitrary coherence
within the doublet) has an intensity that is symmetric under point
reflection through the core center, and therefore illuminates two
diagonal pixels centered on the core image equally; the odd-parity part
of the intensity is exactly the 01--11 coherence term. A \emph{coherent}
superposition
\(\sqrt{1-w_{11}}\,\mathrm{LP}_{01} + e^{i\varphi}\sqrt{w_{11}}\,\mathrm{LP}_{11}\)
adds the interference term
2\(\sqrt{w_{01}w_{11}}\cos\varphi\, f_{01}f_{11}\), which is odd under
that reflection: it brightens one side and darkens the other,
reproducing the observed single-sided lobe (Fig. 4d, computed from the
fiber's mode fields). A single 4-pixel snapshot with a free alignment
offset does not discriminate, since an incoherent donut viewed by a
slightly offset detector reproduces the bare NW/SE ratio; the
discrimination therefore lives in the drift-resolved joint fit over all
time slices of the run. That fit rejects the incoherent-mixture model
(free per-slice alignment offsets) against the coherent model at a raw
\(\Delta\chi^2 \approx 1.9\times10^{3}\), about 3\(\sigma\) after
inflating the errors by the reduced \(\chi^2\) of the
systematics-dominated fit (Fig. S2); the converse does not hold, since a
coherent state with \(\varphi = \pm\pi/2\) also produces a symmetric
pattern. The intermodal phase therefore survives the fiber: each core
carries a spatial qubit. That the interference survives the source
bandwidth at all is informative. The step-index solver with the
effective parameters gives an intermodal group-delay difference of 1.3
ps per meter of fiber, i.e.~50--80 fs for a few-centimeter sample
(spectral fringe period 25--40 nm), so with the installed 10 nm signal
filter alone the cross-term visibility would be limited to 0.6--0.8
(Fig. S3). A refit of the off-center run with the cross-term amplitude
\(V\cos\varphi\) free prefers the fully coherent boundary (best fit
1.0); with errors inflated by the reduced \(\chi^2\),
\(V\cos\varphi < 0.95\) is disfavored at about 4\(\sigma\), which bounds
the visibility (\(V \geq 0.95\)) and the phase
(\(|\varphi| \lesssim 0.3\) rad) simultaneously. The pattern reaches
full contrast because the herald arm selects the idler through the
monochromator, which by energy conservation projects the heralded signal
onto a pump-limited band, \(\approx 3.5\) nm at 800 nm for a
transform-limited 270 fs pump; for this band the predicted visibility
stays above 0.95 for sample lengths up to \(\approx 55\) mm, consistent
with the measured \(V \geq 0.95\). Conversely, the measured visibility
bounds the product of sample length and intermodal dispersion,
\(L\,d(\delta\beta)/d\lambda\), and the asymmetry-vs-bandwidth relation
is an in-situ probe of that product which the classical wavelength-sweep
campaign can check against the predicted fringe period of several tens
of nanometers.

Because the fiber admits only these two modes, every admissible image
belongs to a three-parameter family, and the heralded images can be
inverted for the launch. A joint fit of all time slices of both imaging
runs reconstructs the launch trajectory (Fig. S1a); the fit uses a fixed
detector offset, magnification, and global phase
\(\varphi = \delta\beta L\), together with a free launch coordinate s(t)
per slice and modal content \emph{constrained to the launch-overlap
physics} \(w_{11}(|s|)\). The nominally centered run reveals its own
incoupling drift, \(|s|\): 0.6 \(\rightarrow\) 2--5 \(\mu\)m over the
course of the run, while the deliberately displaced run is stable at
\textbar s\textbar{} = 2.2 \(\pm\) 0.1 \(\mu\)m throughout. The same
trajectory that explains the image shapes also tracks the independently
measured launch coupling efficiency (Fig. S1c; heralded coincidences per
herald), though the observed loss exceeds the transverse-overlap
prediction at some epochs, indicating additional axial/angular drift
components. The hypothesis that the patterns arise from a drifting
\emph{detector} sampling a pure LP\(_{01}\) is rejected at a raw
\(\Delta\chi^{2} \approx 2\times10^{3}\) for one parameter (same
error-inflation convention as above). This mode-constrained
self-calibration requires no additional measurements, since it uses the
experiment's own quantum data.

\includegraphics[width=0.95\linewidth,height=\textheight,keepaspectratio]{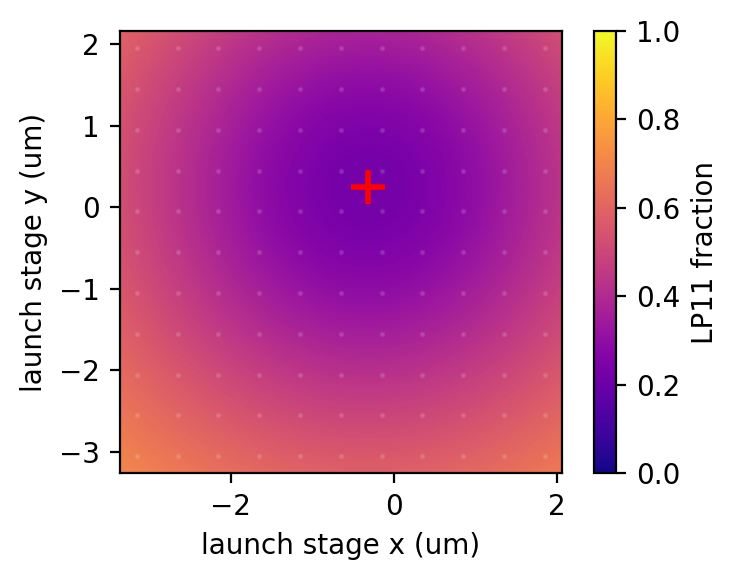}~

\textbf{Fig. 7 \textbar{} Classical calibration of the launch-state
preparation.} Fitted LP11 fraction versus launch-stage position for a CW
800 nm scan, decomposed from camera images of the output facet (dots:
measurement grid; cross: optimum launch). The fraction is minimal at the
optimum, grows radially outward as the launch-overlap model requires,
and reaches about 0.4 at the 2.2 um displacement of the heralded runs,
reproducing the w11 of the photon-level joint fit (Section 3).

The launch-overlap physics behind this preparation is confirmed
classically: a CW 800 nm launch-position scan, decomposed mode-by-mode
from camera images of the output facet, maps the LP\(_{11}\) fraction
over the launch plane (Fig. 7). The fraction is minimal at the optimum
launch, grows radially outward as the overlap model requires, and
reaches \(\approx\) 0.4 at the 2.2 \(\mu\)m displacement used in the
heralded runs, reproducing the \(w_{11}\) found by the photon-level
joint fit. The launch position therefore prepares a known,
phase-coherent LP\(_{01}\)/LP\(_{11}\) content inside the injection
core, calibrated classically and verified photon-by-photon. We now show
that this modal content also acts as a control knob for the lattice.

\subsection{4. State-tunable inter-core
coupling}\label{state-tunable-inter-core-coupling}

Because the LP\(_{11}\) mode couples to the neighboring core
\(\sim 2.4\)\(\times\) more strongly than LP\(_{01}\) (\(\kappa_{01}L\)
= 0.20, \(\kappa_{11}L\) = 0.47, orientation-averaged effective values),
the two mode populations transfer independently
(\(\delta\beta \gg \kappa\)) and the measured crosstalk interpolates
between the two mode-specific values,
\(x = (1-w_{11})\,x_{01} + w_{11}\,x_{11}\) (Fig. 3). It is therefore
set by the launched mode content, that is, by where the input fiber is
positioned; the fiber, the wavelength, and the lattice Hamiltonian are
never touched. The only actuated element is the launch stage, which
prepares the input state upstream of the lattice. The measured
per-photon crosstalk follows this control (Fig. 5b; APD-array optical
crosstalk removed by a pair veto, see Methods): the centered launch and
off-center direction 1 give 3.9--6.4\%, while directions 2 and 3 reach
8.5--13.1\%. The direction dependence also carries mode physics: the two
degenerate LP\(_{11}\) orientations couple differently to the neighbor
(the lobe axis along the inter-core line couples strongly), so the
displacement direction selects the LP\(_{11}\) orientation as well as
its weight, which is why one off-center direction stays near the
centered values. This state-selectable coupling reverses the usual
arrangement: reconfigurable photonic circuits and lattices tune their
Hamiltonian element by
element\textsuperscript{\citeproc{ref-Grafe2016}{32},\citeproc{ref-Carolan2015}{34}},
whereas here a passive, telecom-grade fiber exposes a hopping amplitude
that is tunable through its input state alone, in principle continuously
and demonstrated here over 3.9--13.1\%. The launch stage is of course
also hardware, but it acts before the lattice rather than on it; the
Hamiltonian is fixed at fabrication, and only the state presented to it
is chosen. Immediate uses: (i) quantum walks with launch-selected spread
rate (the walk parameter \(\kappa_{\mathrm{eff}}L\) is set per run; for
intermediate \(w_{11}\) the core populations are the weighted pair of
walks with \(\kappa_{01}\) and \(\kappa_{11}\)); (ii) mode-selective
routing, in which a photon in LP\(_{01}\) traverses a core as an
isolated channel while a photon converted to LP\(_{11}\) couples to the
neighbor bus, a passive switch for mode-multiplexed quantum
interconnects; (iii) crosstalk budgeting in SDM quantum networks, where
the same fiber serves isolation (LP\(_{01}\)) or deliberate
inter-channel operations (LP\(_{11}\)) on demand; (iv) tunable multiport
splitters for boson-sampling-type unitaries in
fiber\textsuperscript{\citeproc{ref-Carine2020}{14},\citeproc{ref-Spring2013}{31}}.
The detector-artifact budget behind these numbers is measured directly.
A pair-\(\Delta t\) analysis (Methods, Fig. 6) identifies a \(\sim 1\)
ns APD-array optical-crosstalk component inside the coincidence window;
a pair veto (rejecting core-3 clicks 0.64--2 ns after an SE click)
removes it and shifts every direction value by only \(\sim 0.07\)
percentage points, so the tunability is not a detector artifact. A
residual core-3 floor (lone clicks; sub-resolution crosstalk vs cladding
light) bounds the values from below: excluding core 3 entirely lowers
all points by 20--24\% (\(\approx 23\%\) on average) without changing
the direction ordering. A prompt SE\(\rightarrow\)C flash component
cannot be timing-separated from genuine core-2 counts; its scale is
bounded by the measured SE\(\rightarrow\)NW/W bump amplitudes at well
below one percentage
point\textsuperscript{\citeproc{ref-Rech2008}{45},\citeproc{ref-Ziarkash2018}{46}}.
\includegraphics[width=0.95\linewidth,height=\textheight,keepaspectratio]{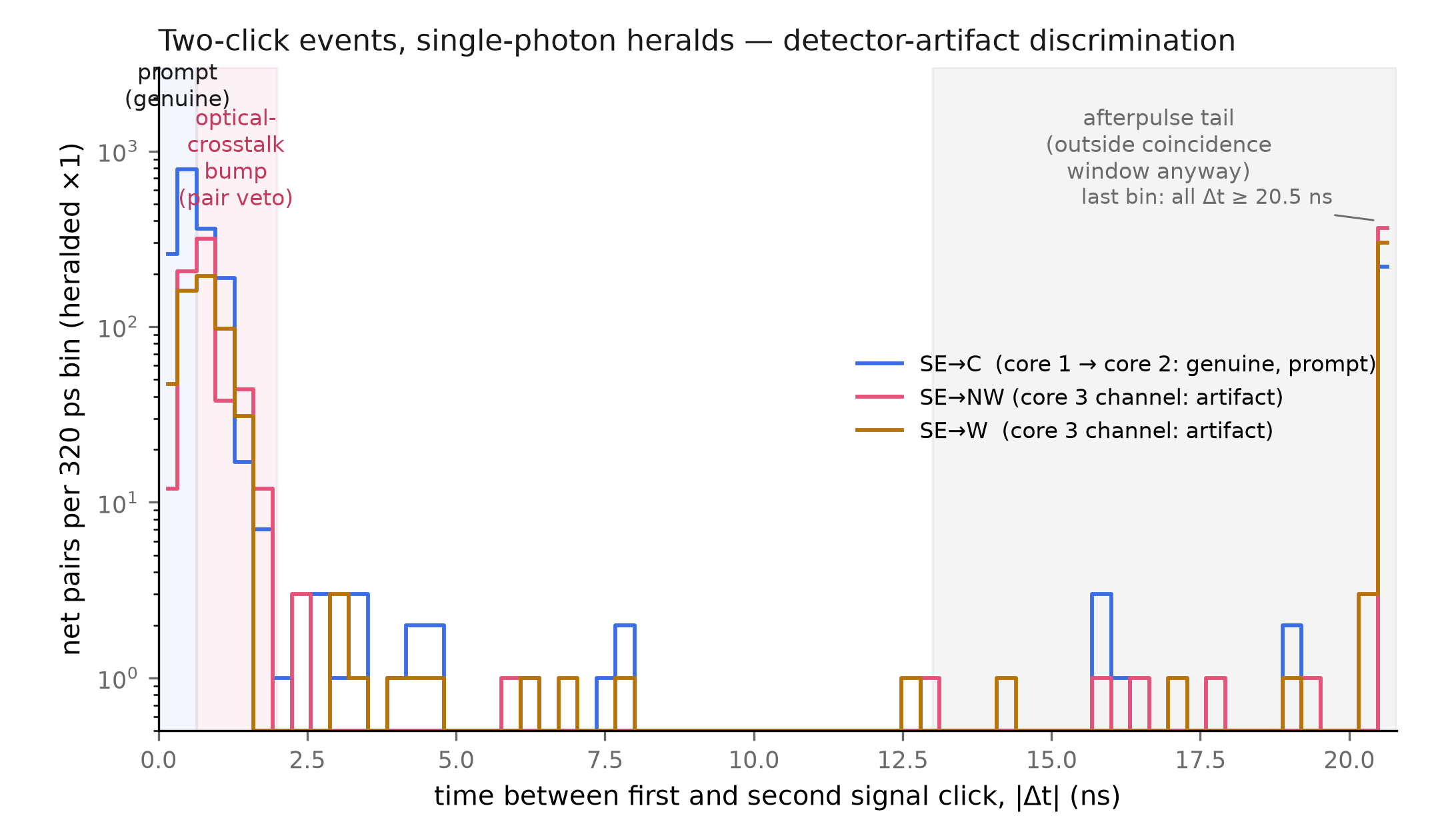}~

\textbf{Fig. 6 \textbar{} Detector-artifact discrimination (Methods).}
Pair time-delay distributions separating prompt genuine coincidences,
the \textasciitilde1 ns APD-array optical-crosstalk bump (removed by the
pair veto), and the afterpulse tail already excluded by the coincidence
window.

The veto is deliberately \emph{not} applied to the photon-number
comparison of Section 5, because it preferentially removes genuine
SE+core-3 split pairs from the heralded \(|2\rangle\) events (a
dead-window bias); in the un-vetoed comparison the APD bump enters
\(x_1\) and \(x_2\) symmetrically and cancels at the quoted precision.
Repeated centered runs span 4.4--6.4\%, making launch stability (Section
3) the dominant systematic. Two datasets are excluded: one direction-3
run with negative net counts (herald-arm fault) and one earlier run in a
different optical configuration; the criteria are net-count sanity and
identical optical configuration.

Every use of this coupling knob for quantum information, however, rests
on one assumption: that the lattice treats one photon and two photons
alike. We verify this next.

\subsection{5. Photon-number characterization: the measured
linear-optics
null}\label{photon-number-characterization-the-measured-linear-optics-null}

\includegraphics[width=0.95\linewidth,height=\textheight,keepaspectratio]{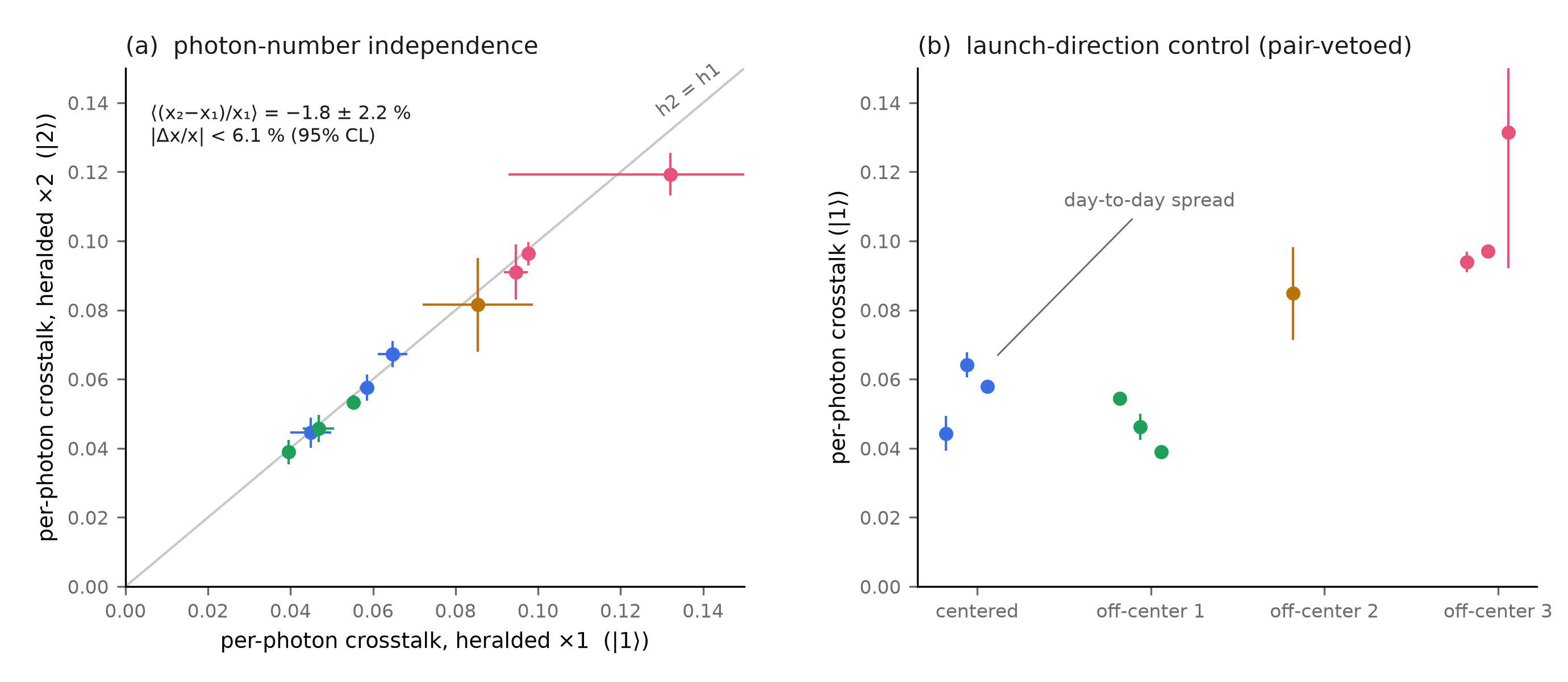}~

\textbf{Fig. 5 \textbar{} Photon-number independence and state-tunable
coupling.} (a) Per-photon inter-core crosstalk of heralded two-photon
events versus single-photon events, one point per run; all points lie on
the h2 = h1 diagonal, bounding any Fock-state dependence to below 6.1\%
(95\% CL). (b) Pair-vetoed per-photon crosstalk by launch configuration:
the launch direction tunes the effective coupling from 3.9\% to 13.1\%.
Colors denote the launch configuration as in (b); panel (a) is un-vetoed
(the APD bump cancels between h1 and h2), panel (b) pair-vetoed.

For a passive linear network, the mean (per-photon) transfer
probabilities are independent of the input photon number: \(|1\rangle\),
\(|2\rangle\), and coherent states in the same wavepacket yield
identical per-photon crosstalk and identical mode images. The per-photon
crosstalk is defined as \(x\) = (net counts in cores 2 and 3)/(net
counts in all cores); at our per-pulse click probabilities same-pixel
pile-up for h2 is negligible. Our data confirm this null in two
independent ways:

\begin{enumerate}
\def\labelenumi{\arabic{enumi}.}
\tightlist
\item
  \textbf{Crosstalk} (core-resolved runs, 0.2--4\(\times10^{6}\) net
  heralded photons per run, more than 10\(^{7}\) in total): the
  per-photon inter-core crosstalk for heralded \(|2\rangle\) equals that
  for \(|1\rangle\) in every run (Fig. 5a). Uncertainties combine
  subtraction-propagated counting statistics (var(net) = raw +
  accidentals) with a per-run drift systematic from the scatter of eight
  time slices; this systematic is assigned to \(x_1\) and \(x_2\)
  independently and is therefore conservative, since the drift is
  largely common mode and cancels in \(x_2 - x_1\) (consistent with
  \(\chi^2/\mathrm{dof} < 1\)). Every run is individually consistent
  with \(x_2 = x_1\) (all \(< 1\sigma\)), and the ensemble weighted mean
  is \(\langle x_2 - x_1\rangle = -0.0012 \pm 0.0014\)
  (\(\chi^{2} = 0.9/9\)),
  i.e.~\(\langle(x_2-x_1)/x_1\rangle = -1.8 \pm 2.2\,\%\). \textbf{Any
  photon-number dependence of the per-photon crosstalk is bounded to
  \(|\Delta x/x|\) \textless{} 6.1\% at 95\% confidence}, across launch
  conditions spanning 3.9--13.1\% absolute crosstalk.
\item
  \textbf{Mode distribution} (imaging runs): the heralded pixel
  distributions of \(|1\rangle\) and \(|2\rangle\) agree to \textless1\%
  for both launches (Fig. 4b,c,e,f); for example, off-center 50.2/47.1\%
  (\(|1\rangle\)) vs 49.9/47.2\% (\(|2\rangle\)).
\end{enumerate}

The null is nontrivial to measure, because coincidence-based event
selection breaks it \emph{spuriously}. If accidentals are not
subtracted, or if analyses select events where all n heralded photons
appear in one core, the apparent \(|2\rangle\) crosstalk is strongly
suppressed relative to \(|1\rangle\), because for independent photons
the probability that \emph{both} leak scales quadratically. Our
calibrated coupled-mode model (end-of-chain injection; \(\kappa_{01}L\)
= 0.20, \(\kappa_{11}L\) = 0.47, anchored on the \(|1\rangle\) data
alone) quantifies the effect: with the event-selection filter, the
apparent \(|2\rangle\) crosstalk drops from 10\% to 1\% (off-center
launch) with \emph{no change in physics}. Exact Fock-space propagation
(Perceval SLOS) and matrix-permanent evaluation (The Walrus) agree with
the closed-form multinomial statistics to \textless{}\(10^{-13}\) over
all two-photon patterns; no genuine multiphoton interference is possible
for a single occupied input wavepacket, which makes this dataset the
validated baseline against which interference-driven deviations
(e.g.~two-photon inputs distributed over two cores) must be measured.
(Fig. S4.)

\subsection{6. Discussion and outlook}\label{discussion-and-outlook}

Heralded \(|2\rangle\) events with both photons detected provide the
first correlated observables: \(\sim 2.3\times10^{3}\) split pairs
(photons in two different cores) across the series, and pixel-pair
patterns within the launch core in the imaging runs. The split-pair
rates are consistent with independent-photon statistics, as the null
requires. In the imaging runs, pairs involving the NW pixel (the dimmest
in that configuration) appear \(\sim 10\)\(\times\) above the
independent expectation; their pair time-delay distribution places them
in the \(\sim 1\) ns APD-array optical-crosstalk bump rather than in the
prompt window, which identifies them as detector artifacts (avalanche
flash) and not as spatial correlation (Methods, Fig. 6). These
correlations are the first step toward multiphoton mode analysis on this
platform.

The platform combines three measured properties: a spatial qubit per
core with phase-preserving propagation, the resource that quantum-walk
and boson-sampling proposals assume; launch-selectable coupling
(3.9--13.1\%) at fixed fiber and wavelength; and a calibrated
multiphoton baseline validated by three independent engines. The next
characterization steps are a heralded \(g^{(2)}(0)\) of each Fock class
at the fiber output and two-photon interference across cores. The
measured null defines the next experiment: injecting the two photons of
\(|2\rangle\) into \emph{two different cores} (or two modes of one core)
enables genuine Hong--Ou--Mandel-type quantum-walk
interference\textsuperscript{\citeproc{ref-Bromberg2009}{27},\citeproc{ref-Peruzzo2010}{28}},
and a deviation from our multinomial baseline is then unambiguously
quantum; two-photon HOM interference across MCF cores has just been
demonstrated for inter-core-skew
metrology\textsuperscript{\citeproc{ref-Skew2026}{47}}, and
photon-number-dependent lattice dynamics have so far been reached only
through classical emulation\textsuperscript{\citeproc{ref-Hui2025}{48}}.
The intermodal phase at nanometer-scale bandwidths dephases after a few
meters of fiber (\(\sim 1.3\) ps/m walk-off), so mode-encoded
superpositions are a short-reach resource, while the which-core degree
of freedom retains its kilometer scaling. The mode-constrained
self-calibration (Section 3) transfers to any few-mode quantum platform.
Beyond quantum information, a calibrated Fock-state network with
number-resolved readout is also the setting proposed for foundational
tests of measurement statistics: alternative statistical models of
photon-number measurement make beam-splitter predictions that differ
from quantum mechanics\textsuperscript{\citeproc{ref-Powers2025}{49}},
and our validated multinomial baseline supplies exactly the reference
such a discrimination requires. The state-tunable coupling opens a
further direction: with two modes per core, the lattice is naturally a
two-leg ladder, and the preserved intermodal phase \(\varphi\) plays the
role of a synthetic flux through its plaquettes, thermally tunable at
\(\sim 0.03\) rad/K per centimeter of fiber (from
\(L\,d(\delta\beta)/dT\) with the silica thermo-optic coefficient),
which suggests fiber-based quantum simulation of ladder models and
flux-dependent (Aharonov--Bohm-cage-like) transport with genuine quantum
light. On the network side this makes the mode-selective router of
Section 4 concrete: a single mode converter switches a photon between
``transmit'' (LP\(_{01}\), isolated) and ``distribute'' (LP\(_{11}\),
coupled to the neighbor bus) in deployed-compatible fiber. The readout
scales to all 39 cores with commercially available linear SPAD
arrays\textsuperscript{\citeproc{ref-PiImaging2026}{50}}, and
single-detector photon-number resolution is advancing
independently\textsuperscript{\citeproc{ref-Schapeler2024}{51}}. Tunable
multiport unitaries in
fiber\textsuperscript{\citeproc{ref-Carine2020}{14},\citeproc{ref-Reck1994}{52}}
then put boson-sampling-style
experiments\textsuperscript{\citeproc{ref-Spring2013}{31}} in reach. At
1550 nm, SNSPD arrays offer the same scaling, and the platform is
compatible with fiber-coupled single-photon
emitters\textsuperscript{\citeproc{ref-Bommer2011}{53}--\citeproc{ref-Kilinc2021}{55}}.
Phase-preserving frequency conversion at the single-photon
level\textsuperscript{\citeproc{ref-Cheng2015}{56}} indicates that such
phase-encoded states can also be wavelength-shifted without losing the
interferometric information. The measured \(\varphi\) =
\(\delta\beta L\) yields the intermodal beat length of the deployed
fiber from quantum data alone.

\subsection{Methods}\label{methods}

\textbf{Fiber:} the multicore ring fiber
of\textsuperscript{\citeproc{ref-Nguyen2020}{5}} (39 cores on a C-shaped
arc; design parameters therein); effective step-index description
\(a_{\mathrm{eff}}\) = 4.5 \(\mu\)m, \(V_{\mathrm{eff}}\)(800 nm) = 3.57
(LP\(_{01}\)+LP\(_{11}\)), fitted to the mode images; couplings fitted
(below). \textbf{Source \& herald:} type-II SPDC/BBO; 249,733 Hz sync;
herald arm = cascaded FBS onto 4
APDs\textsuperscript{\citeproc{ref-Thomay2017}{39}}; photon-number
statistics modeling follows\textsuperscript{\citeproc{ref-Xu2024}{41}};
herald multiplicity distribution (single centered run):
1.16\(\times10^{9}\) / 6.9\(\times10^{7}\) / 2.7\(\times10^{6}\) /
5.1\(\times10^{4}\) pulses for m = 1--4. \textbf{Acquisition:} PicoQuant
MultiHarp, T3, 320 ps resolution; \(\sim 113\) GB raw time tags;
datasets and analysis code archived (see Data availability).
\textbf{Coincidence pipeline} (independent implementation, archived with
the data): pulse-index grouping; \(\Delta_0\) = 4.16 ns calibrated from
unconditional dtime histograms (all four signal channels share one peak
bin, confirming the absence of inter-core delay skew); dtime windows 1
and 3 ns (results window-insensitive over 0.5--3 ns in the pulse-matched
pass); accidentals from pulse-(k+1) matching subtracted from every
counter; core\(\leftrightarrow\)channel map validated in-data (centered
launch dominates the core-1 channel). \textbf{Imaging analysis:}
time-sliced (8 slices of 3 h per 24 h run) heralded per-pixel counts;
joint launch-drift fit as in Section 3 (fixed detector offset D,
magnification M, intermodal phase \(\varphi\), array roll \(\psi\);
per-slice launch s; \(w_{11}\)(\textbar s\textbar) from overlap
integrals). \textbf{Classical launch calibration (Fig. 7):} CW 800 nm
light launched through the same 780HP fiber and piezo stage, launch
position raster-scanned; the imaged output facet is decomposed per
position into the two guided-mode intensities, giving the fitted
LP\(_{11}\) fraction over the launch plane. \textbf{Simulation:}
step-index LP solver (weakly guiding LP formalism and characteristic
equation follow\textsuperscript{\citeproc{ref-Ghatak1998}{57}}, Ch. 8);
coupled-mode Hamiltonian on
\(\{\mathrm{core}\}\otimes\{\mathrm{LP}_{01},\mathrm{LP}_{11}\}\), open
chain with end-of-chain injection, \(\delta\beta\) \(\gg\) \(\kappa\);
\(\kappa\) calibrated on \(|1\rangle\) anchors (centered 4.6\%,
off-center high-coupling direction 10\%): \(\kappa_{01}L\) = 0.197,
\(\kappa_{11}L\) = 0.469 (the data fix only the products \(\kappa L\));
unitarity 3\(\times10^{-13}\); \textbf{calibration caveat:}
neighbor-count ratios fold in unequal per-core pixel-collection
efficiencies (arc-on-grid geometry; core 3 straddles two pixels), so the
\(\kappa L\) values carry a collection systematic; the h1/h2 null is
unaffected (efficiencies cancel). APD-array optical crosstalk and
afterpulsing follow the known SPAD
phenomenology\textsuperscript{\citeproc{ref-Rech2008}{45},\citeproc{ref-Ziarkash2018}{46},\citeproc{ref-Cova1996}{58}}
and are handled by the pair-\(\Delta t\) veto (Fig. 6); Fock statistics
closed-form, cross-validated with Perceval 1.2.3 (SLOS backend) and The
Walrus
0.22\textsuperscript{\citeproc{ref-Perceval2022}{59}--\citeproc{ref-Gupt2019}{61}}
to \textless{} 10\(^{-13}\). \textbf{Provenance:} November 2025 series
measured by T.T. and C.W.; classical reference campaign remeasured June
2026.

\textbf{Funding:} National Science Foundation, EAGER award 2533850.
\textbf{Acknowledgments:} We thank Corning Incorporated for the
multicore ring fiber, Aramais Zakharian for fruitful discussions, Boh
Roffin and Karl Kock for their support, and OptoElectronic Components
for the loan of the 3\(\times 3\) APD array and the MultiHarp.
\textbf{Data availability:} The analysis code, the coupled-mode/Fock
simulation package, and per-run derived data (coincidence counters,
modal weights, fit results) are archived on Zenodo at
https://doi.org/10.5281/zenodo.23102388. The raw time-tag records
(\textgreater{}\(10^{10}\) events, PicoQuant PTU) are archived and
available on reasonable request.

\textbf{Supplementary material:} Supplementary figures S1 to S4 will be
provided with the journal version.

\subsection{References}\label{references}

\phantomsection\label{refs}

\end{document}